\documentclass[psfig]{aa}
\input epsfig.sty

\begin{document}

   \thesaurus{03(13.25.2 -- 11.09.1 -- 11.19.1)}
  \title{BeppoSAX observations of Narrow-Line Seyfert 1 galaxies: II. 
        Ionized iron features in Arakelian~564}

  \author{A. Comastri 
              \inst{1}
    \and     G.M. Stirpe 
              \inst{1}
     \and    C. Vignali 
              \inst{1,2}
     \and    W.N. Brandt       
               \inst{3}
     \and    K.M. Leighly
               \inst{4} \fnmsep\thanks{Current Address : Department of 
 Physics and Astronomy, The University of Oklahoma, 440 W. Brooks St., 
 Norman OK 73019, USA}
     \and    F. Fiore 
               \inst{5,6,7}
     \and    M. Guainazzi
               \inst{8} 
      \and    G. Matt
               \inst{9}
      \and   F. Nicastro 
               \inst{7}
      \and   E.M. Puchnarewicz
               \inst{10}
      \and   A. Siemiginowska 
               \inst{7}
} 
  \offprints{A. Comastri (comastri@bo.astro.it)}

  \institute {Osservatorio Astronomico di Bologna, Via Ranzani 1, I--40127
              Bologna, Italy
  \and       Dipartimento di Astronomia, Universit\`a di Bologna,
             Via Ranzani 1, I--40127  Bologna, Italy
    \and  Dept. of Astronomy and Astrophysics, The Pennsylvania
         State University, 525 Davey Lab, University Park, PA 16802, USA
    \and Department of Astronomy, Columbia University, 538 West 120th Street,
          New York, NY 10027, USA
    \and  Osservatorio Astronomico di Roma, Via Frascati 33, I--00044
           Monteporzio--Catone, Italy
    \and  SAX--Science Data Center, Nuova Telespazio, Via Corcolle 19,
           I--00131 Roma, Italy
    \and Harvard--Smithsonian Center for Astrophysics, 60 Garden St.,
            Cambridge, MA 02138, USA
    \and  XMM-Newton SOC, VILSPA - ESA, Apartado 50727, E-28080 Madrid 
          Spain  
    \and  Dipartimento di Fisica ``E. Amaldi", Universit\`a degli Studi
          ``Roma Tre", Via della Vasca Navale 84, I--00146 Roma, Italy
    \and  Mullard Space Science Laboratory, University College London,
          Holmbury St. Mary, Dorking, Surrey RH5 6NT, UK
}

  \date{Received 21 July 2000/ Accepted 23 October 2000}

\titlerunning{BeppoSAX observations of NLS1: Ark 564}
\authorrunning{A. Comastri et al.}

  \maketitle

\begin{abstract}

The BeppoSAX observations of the bright Narrow--Line Seyfert 1 galaxy 
Ark~564 are presented along with a high quality optical spectrum
taken at the 1.5m telescope at La Silla.
The 0.1--10 keV X--ray spectrum is characterized 
by a strong soft component which is best described by blackbody--like 
emission with a temperature of $\sim$ 160 eV. At higher energies a steep 
($\Gamma \simeq$ 2.4) power--law tail is present. 
There is evidence of an ionized reflector 
in the form of an iron line and edge. We do not find significant evidence 
of soft X--ray features if the spectrum is modelled with a two component 
continuum. The optical and X--ray spectral properties support the 
hypothesis of a high accretion rate onto a low mass black hole.


 \end{abstract}

\section{Introduction}

In recent years sensitive ROSAT and ASCA observations 
have revealed that a class of objects classified on the basis of their
optical spectra as Narrow--Line Seyfert 1 galaxies (hereinafter NLS1)
exhibits peculiar and rather extreme properties in the X--ray band.
Large amplitude X-ray variability with timescales as fast as a few minutes
(Boller et al. 1997) and extremely soft X--ray spectra in the ROSAT 
band are common among these objects (Boller et al. 1996). 
ASCA observations have extended the ROSAT results to higher energies.
On average, NLS1 galaxies have steeper X--ray spectra 
(Brandt et al. 1997) and higher variability amplitude
variance (Fiore et al. 1998; Leighly 1999a; Turner et al. 1999a) 
compared to Seyfert 1s with broad optical lines. 
The anti--correlation between the X--ray spectral slope 
and the H$\beta$ FWHM is now a well--established observational property
tested across a wide range of X--ray luminosities (Laor et al. 1997),   
suggesting that NLS1 represent an extreme of Seyfert activity, possibly
linked to an extreme value of a fundamental physical parameter.

The ASCA observation of REJ 1034+393 can be considered 
a breakthrough in the understanding of NLS1, because it was the 
first object of this class observed above 2 keV. The 0.5--10 keV 
spectrum is dominated by a strong soft component below $\sim$
2 keV; a steep ($\Gamma \simeq$ 2.6) power law tail at higher
energies and an ionized iron K$\alpha$ line at 6.7 keV
(although detected only at the 2$\sigma$ level) 
are also present (Pounds et al. 1995). 
Ionized iron lines have been discovered in a few other objects observed
by BeppoSAX (Comastri et al. 1998, paper I) and ASCA 
(Turner et al. 1998; Leighly 1999b),   
suggesting that the fundamental parameter 
could be a particularly high accretion rate (Pounds et al. 1995; 
Laor et al. 1997). 

We have started an observational programme to study a sizeable sample of NLS1 
with BeppoSAX with the aim of investigating the broad band X--ray spectral
and variability properties of these objects. We are taking advantage 
of the increased effective area at high energies, especially above 5 keV,
where the MECS detectors provide the best opportunity to
investigate the iron line properties.

Arakelian 564 ($z$ = 0.0247, $V$ = 14.16) is an ideal candidate for such an 
investigation being the brightest NLS1 in the 2--10 keV band. 
Previous ROSAT and ASCA observations revealed a complex X--ray spectrum 
with a few absorption/emission features around 1 keV 
(Brandt et al. 1994; Leighly 1999b; Vaughan et al. 1999a; 
Turner et al. 1999b, hereinafter TGN99). 
More recently, an RXTE/PCA observation (Vaughan et al. 1999b, hereinafter V99) 
provided the first spectrum of this object up to 20 keV. 
The most remarkable result is the presence of a deep ionized edge 
at $\sim$ 8.5 keV interpreted as reflection from an ionized disc.
In the present paper we report the results of 3 BeppoSAX
observations of Ark 564 together with a high--quality optical 
spectrum obtained at La Silla one year before the first X--ray observation.  

\section{The optical spectrum}

\subsection{Observations and reduction}

Optical spectra of Ark 564 were obtained in photometric conditions at the
ESO 1.52m telescope on 1996 October 1, using the Boller \& Chivens
spectrograph with a 127mm camera and a Loral/Lesser thinned CCD with
$2048\times2048$ pixels.  The pixel size is $15\mu\hbox{m}$, and the
projected scale on the detector 0.82~arcsec~pixel$^{-1}$.  The grating used
has 600~grooves~mm$^{-1}$.  The spectra were obtained through a 2-arcsec
wide slit, at a resolution of 4.6~\AA.  Two integrations of 1200 seconds
each were obtained, with mid-point airmasses of 2.05 and 1.97.

\begin{figure*}
\epsfig{file=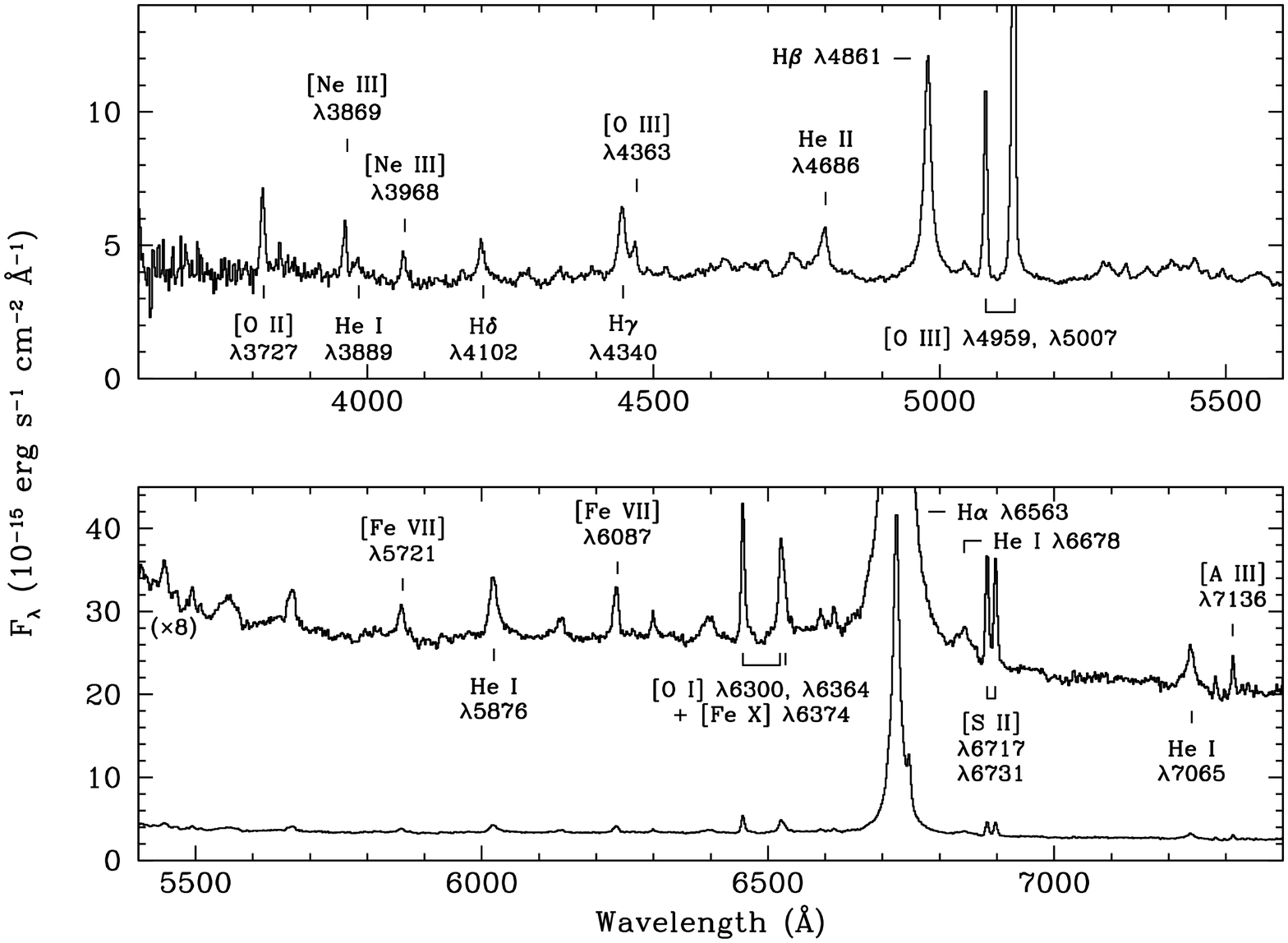, height=12.cm, angle=-90, silent=}
\caption{The optical spectrum of Ark 564 obtained at ESO in 1996 October. The
spectrum has been split in two segments for viewing purposes.  The bottom
panel also shows the red segment scaled up by a factor 8, in order to
evidence the weaker lines.  All lines which are not explicitly labelled on
the figure can be identified with Fe II features.  The high frequency
structure around the [A~III]$\lambda$7136 line is caused by residuals from
the removal of the atmospheric O$_2$ A-band.  The wavelength scale is in the
rest system of the observer ($z$ = 0.025).}
\end{figure*}

Standard techniques were used to reduce the spectra, using the NOAO IRAF
package.  The reduction included the correction of the atmospheric
absorption bands of O$_2$ and H$_2$O, obtained by interpolating and
normalizing the continuum of the spectrophotometric standards.  A short
exposure of Ark 564 obtained during the same night through an 8~arcsec slit
was used to correct the narrow-slit spectra for light losses and
differential refraction.  This was achieved by fitting a low-order
polynomial to the ratio between each narrow-slit spectrum and the broad-slit
one, and dividing each narrow-slit spectrum by the corresponding fit.
Despite this correction, the low culmination point of the target with
respect to the observing site still leaves a considerable uncertainty on the
spectrophotometric quality of the data, particularly at
$\lambda\le4500$~\AA.  The two narrow-slit spectra were averaged and yielded
the final spectrum shown in Fig.~1.  The calibration curves
obtained during the night and the individual spectra coincide to within
$<$5\%, thus we consider the spectrophotometric quality of the red part of the
spectrum to be of this order.

\subsection{Properties of the optical spectrum}

While the widths of the permitted lines fully justify the classification of
Ark 564 as a NLS1, other characteristics of its optical spectrum are not as
extreme as usually observed among NLS1.  In particular, the source displays
a strong and rich forbidden line spectrum, while the Fe~II lines have
strengths more similar to those of normal Seyfert 1 nuclei rather than the
high values displayed by many NLS1: the ratio between the 
Fe~II~$\lambda$6250 blend and H$\beta$ is 0.6, only slightly higher than the 
typical value for classic Seyfert 1 spectra (Joly 1988).

The optical continuum is flat and likely to include a significant stellar
component.  For this reason, and because of the uncertainty in the
calibration of the blue part of the spectrum, no attempt was made to fit a
continuum across the entire spectrum.  Instead, local straight-line continua
were fitted under the main lines in order to measure their fluxes.

The fluxes of H$\beta$ and [O~III]$\lambda5007$ were measured after
subtraction of the strongest blending lines with suitable templates.  For
Fe~II~$\lambda4924$ and Fe~II~$\lambda5018$ we used the profile of H$\alpha$
scaled by a factor 0.015, under the reasonable assumption that the lines in
multiplet 42 are of equal intensity.  To subtract [O~III]$\lambda4959$ we
used the deblended profile of [O~III]$\lambda5007$ scaled by one third. The 
intensity of the narrow components of the Balmer lines, and the [N~II] lines 
which contaminate H$\alpha$ were estimated by subtracting the [O~III] template, 
scaled in order to obtain a smooth residual profile. The fluxes, widths and 
equivalent widths of the individual lines and components thus obtained are 
listed in Table~1.

\begin{table}
\caption{Integrated fluxes, FWHMs, and equivalent widths 
of the main optical emission features}

\vglue0.3truecm
{\hfill\begin{tabular}{l c c c}
 Line$^a$  & Flux$^b$ & FWHM$^c$ & EW$^d$ \\
&&\\
\hline  
 $[O II] \lambda$ 3727   & 2.6 &  630 &   6 \\
 $[Ne III] \lambda$ 3869  &  1.7 &  600  &  4 \\
 He I $\lambda$ 3889 + H$\zeta$   & 1.1  & 1300 &  3 \\
 $[Ne III] \lambda$3968 + H$\eta$ & 1.4 &   900 &  4 \\
 H$\delta$ + $[S II]\lambda$4068,$\lambda$4076 &  2.6 & 1400 &  7 \\
 H$\gamma$ broad  &  5.0 & 1030 &  14 \\ 
 H$\gamma$ narrow  &  0.4 &  400  &  1 \\
 $[O III]\lambda$4363                   &  0.7 &  400  &  2 \\
 Fe II $\lambda$4570 blend             & 11.9 & ... & 32 \\
 He II $\lambda$4686                    & 4.5 &  1100 & 12 \\
 H$\beta$ broad                         & 17.0 &  960 &  45 \\
 H$\beta$ narrow                        & 1.3 &  400 &  3 \\
 $[O III] \lambda$4959                   & 5.4 &  400 &  16 \\
$[O III] \lambda$5007                   & 18.0 & 400 &  49 \\
Fe II $\lambda$5250 blend            & 15.1 & ... &  43 \\
 $[Fe VII]\lambda$5721                 & 0.5 & 630 &  2 \\
He I $\lambda$5876                    & 2.5 & 1040  & 7 \\
$[Fe VII] \lambda$6087                & 0.8 &  500 &   3 \\
$[O I]\lambda$6300                  & 1.7  &  330 &   5 \\
$[O I]\lambda$6364                  &  0.7  & 340 &   2 \\
$[Fe X]\lambda$6374                &  2.6 &   700 &   8 \\
$[N II]\lambda$6548                 &  1.2 &   400 &   4 \\
H$\alpha$ broad                    & 89.3 &  700 &  300 \\
H$\alpha$ narrow                  &   3.6  &  400  &   12 \\
$[N II]\lambda$6584                &     3.6 &   400   &  12 \\
He I $\lambda$6678                 &    1.3  &   1520  &    4 \\
$[S II]\lambda$6717                 &    1.3  &        300 & 4 \\ 
$[S II]\lambda$6731                  &  1.2    &      300  & 4  \\
He I $\lambda$7065                    & 1.1     &     790   &   4 \\
$[A III]\lambda$7136                   & 0.4     &     250   &   2 \\
\hline 

\end{tabular}\hfill}

$^a$ All wavelengths are in the rest frame \\
$^b$ Units of $10^{-14}$ erg cm$^{-2}$ s$^{-1}$ \\
$^c$ km s$^{-1}$ \\
$^d$ $\AA$ 

\end{table}

\section{BeppoSAX Data reduction}

Arakelian 564 was observed with the BeppoSAX (Boella et al. 1997a) 
Narrow Field Instruments: LECS (0.1--10 keV; Parmar et al. 1997),
MECS (1.3--10 keV; Boella et al. 1997b), HPGSPC (4--60 keV; Manzo et al. 1997)
and PDS (13--200 keV; Frontera et al. 1997).
We will focus on the results obtained from the imaging gas scintillation 
proportional counters LECS and MECS. The source was not detected in the
HPGSPC nor in the PDS (but see below).
The observations were performed with 2 active MECS units (after the failure of
the MECS1 unit on 1997 May 6) in 3 different periods.
The exposure times and background--subtracted count rates in the 3 
observations are reported in Table 2.
The shorter LECS exposure is caused by the 
instrument being switched off over the illuminated Earth,
 while the useful on--source time
for the PDS is about half of the MECS exposure because the PDS monitors
the source and background alternately.

Standard data reduction techniques were employed following the 
prescriptions given by Fiore et al. (1999).
LECS and MECS spectra and light curves were extracted from
regions with radii of 8 arcmin and 4 arcmin respectively, in order to maximize
the accumulated counts at both low and high energies.
Background spectra were extracted from high Galactic latitude
``blank"  fields and also locally from an annular region around the target.
The mean local background is consistent with that of the blank field 
both in the LECS and in the MECS. 
The PDS data were reduced using the variable rise time threshold technique
to reject particle background.
The source is not detected by the PDS in any of the 3 periods.
After combining the observations, a weak signal is present in the first 
channels
(13--30 keV) with a formal detection significance of $\sim$ 2.8$\sigma$.
Taking into account the  systematic uncertainty in the 
PDS background subtraction (Guainazzi and Matteuzzi 1997), 
the high energy flux of Ark~564 is only about 2$\sigma$ above the background.

The 3 LECS and MECS data sets were analyzed separately in order to investigate
possible variations between the 3 observing periods and within each
observation.

\begin{figure}
\epsfig{file=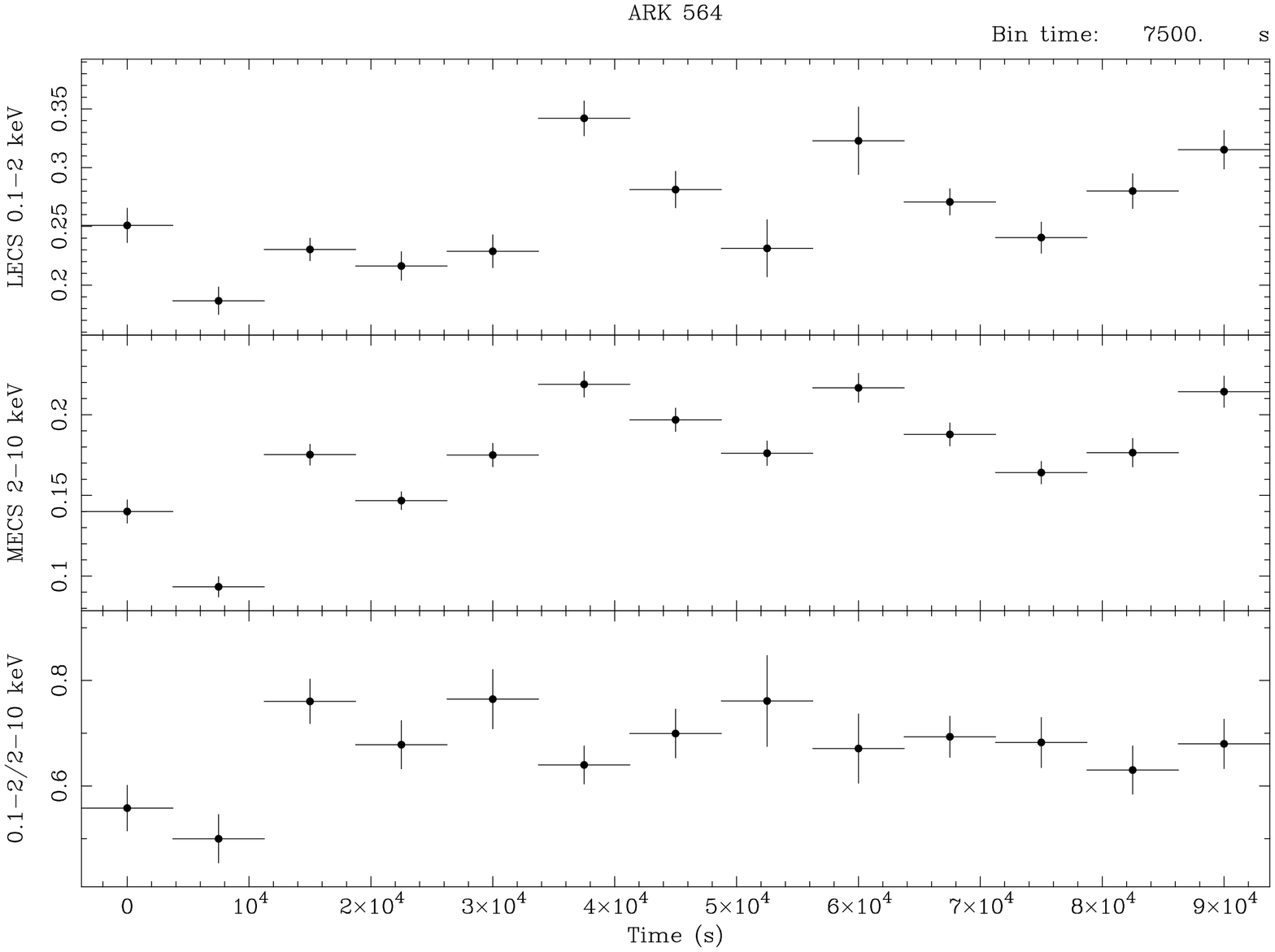, height=7.cm, angle=-90, silent=}
\caption{The LECS (0.1-2 keV, upper panel),  MECS (2-10 keV, middle panel)
and softness ratio (lower panel) light curves for the June 1998 observation.}
\end{figure}

\begin{table}
\caption{BeppoSAX observation log}

\vglue0.3truecm
{\hfill\begin{tabular}{l c c c c}
  Dates & \multicolumn{2}{c}{Exposures (s)} & \multicolumn{2}{c}
{Count Rate (cts/s)}  \\ 
   & LECS  & MECS &  LECS  & MECS   \\
&&\\
\hline
14/11/97 & 12574 & 26196 & 0.254$\pm$0.005 & 0.170$\pm$0.003  \\ 
\hline
12/06/98 & 20209 & 46765 & 0.250$\pm$0.004 & 0.169$\pm$0.002  \\
\hline 
22/11/98 & 11963 & 29154 & 0.255$\pm$0.005 & 0.186$\pm$0.003  \\
&&\\
\hline

\end{tabular}\hfill}

Count rates are in the 0.2--2 keV and 2--10 keV bands for the LECS and 
MECS, respectively.

\end{table}

The source is clearly variable on timescales of a few thousands of seconds
in all observations. In Fig.~2 the LECS and MECS light curves are 
shown for the longest exposure (June 1998). The light curves
of June 1997 and November 1998 are similar in amplitude and 
timescale of variability. 
Significant variability up to a factor 2 is present in both 
energy ranges. 
There is some evidence of a harder spectral state at the 
beginning of the June 1998 observation (Fig.~2). 
As the hardness ratio technique would be insensitive 
to variations in the iron K$\alpha$ flux, we have compared  
the 5--7 keV and 6--7 keV count rates for each observation.
The amplitude of variations in these two narrow bands
is of the order of 10--15 \%, indicating the lack of significant 
flux variability.

Spectral variability could be present at some level, but 
it does not appear to be relevant since for most of the
observing periods, including the November 1997 and November 1998 observations,
the hardness ratio light curve is rather constant. Therefore  
the spectral analysis was carried out using all the 
available data. The resulting exposure times are 
44.7 and 102.1 ks in the LECS and MECS, respectively.

\section{Spectral analysis}

\subsection{The high--energy spectrum}

Spectral analyses were performed using the XSPEC 10 software package and
the latest release of the response matrices.
The MECS spectrum was rebinned in order to obtain at least 30 counts per
energy channel and to sample the energy resolution of the
detector with 3 channels at all energies where possible using 
the appropriate grouping file (see Fiore et al. 1999). The residuals of a 
power--law fit to the 2--10 keV spectrum
indicate the presence of line--like excess emission at 6--7 keV
and of an edge--like structure at energies greater than 8 keV (Fig.~3).
 
\begin{figure}
\epsfig{file=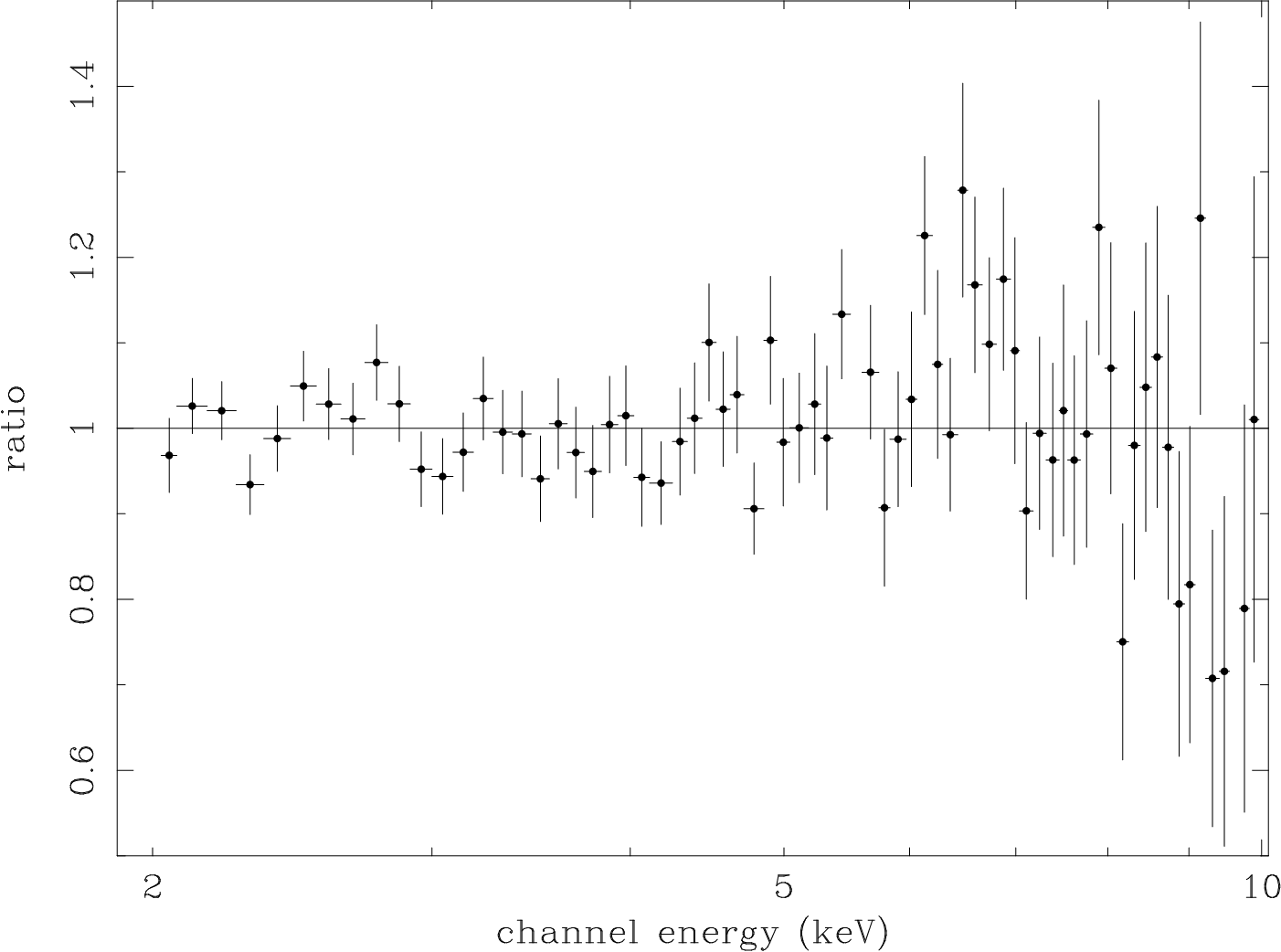, height=7.cm, angle=-90, silent=}
\caption{The residuals of a single power--law fit to the 2--10 keV MECS 
spectrum.}
\end{figure}

The MECS data are best fitted (Table~3) with a rather steep power law 
($\Gamma \simeq$ 2.4) plus a narrow iron emission line at $\sim$ 6.8 keV 
and an absorption edge at $\sim $ 9.5 keV in the source rest--frame 
($z$ = 0.025). The addition of both features improves the fit at the 
99.99 \% level according to an $F$--test. There is no further improvement in 
the fit if the line width is left free to vary below 
the BeppoSAX MECS energy resolution at 6 keV (about 500 eV FWHM).
The $F$ statistic was applied to test the significance
of the individual components. The results indicate that the edge
and the line are required at 99.96 \% and  99\% confidence respectively.
A contour plot of the line versus edge energies clearly 
suggests that they originate from highly ionized gas (Fig.~4).
Indeed the ionization state implied by the best--fit edge energy 
is greater than {\tt Fe XXIII} at the 99\% confidence level.
A similar constraint is obtained from the energy and equivalent width 
of the iron line (see Matt et al. 1993).
The observed features could be due either to gas along the 
line of sight (a warm/hot absorber), or arise in the surface layers 
of a highly ionized optically thick accretion disc (a warm/hot reflector).
A good quality E $>$ 10 keV spectrum or a tight upper limit on the high--energy
flux would distinguish between the two possibilities.
Unfortunately, the marginal detection in the PDS energy range   
is consistent with both the extrapolation of the best fit 
2--10 keV spectrum and with a flattening due to the onset of a 
reflection component as long as the relative normalization between 
the reflected and direct flux is $<$ 3.
Given that in both cases the imprints of ionized absorption/reflection
would also affect the spectrum below 2 keV, we postpone the discussion of these
models to the next section.
The best fit 2--10 keV flux of 1.45 $\times$ 10$^{-11}$ erg cm$^{-2}$ s$^{-1}$
corresponds to a luminosity of $\simeq$ 4 $\times$ 10$^{43}$ erg s$^{-1}$.

\begin{table*}
\caption{MECS fits in the 2--10 keV energy range}

\vglue0.3truecm
{\hfill\begin{tabular}{l l l l l l}
 $\Gamma$ & $E_{line} (keV) $ & EW (eV) & $E_{edge} (keV) $ & $\tau_{edge}$ &
$\chi^2$/d.o.f. \\
\hline
 2.40$\pm$0.04  & ... & ... & ... & ... & 87.4/67 \\
&&\\
 2.42$\pm$0.04  & 6.78$^{+0.19}_{-0.25}$  & 135$\pm$65 & ... & ... &
 75.8/65 \\
&&\\
 2.37$\pm$0.04  & ... & ... & 9.49$^{+0.29}_{-0.41}$ & 0.7$^{+0.67}_{-0.34}$
 & 68.8/65  \\
&&\\
 2.40$\pm$0.04  & 6.78$^{+0.20}_{-0.28}$ & 115$\pm$65 & 9.52$^{+0.30}_{-0.40}$
 & 0.67$^{+0.71}_{-0.35}$  & 60.5/63 \\
&&\\
\hline
\end{tabular}\hfill}


\end{table*}

\begin{figure}
\epsfig{file=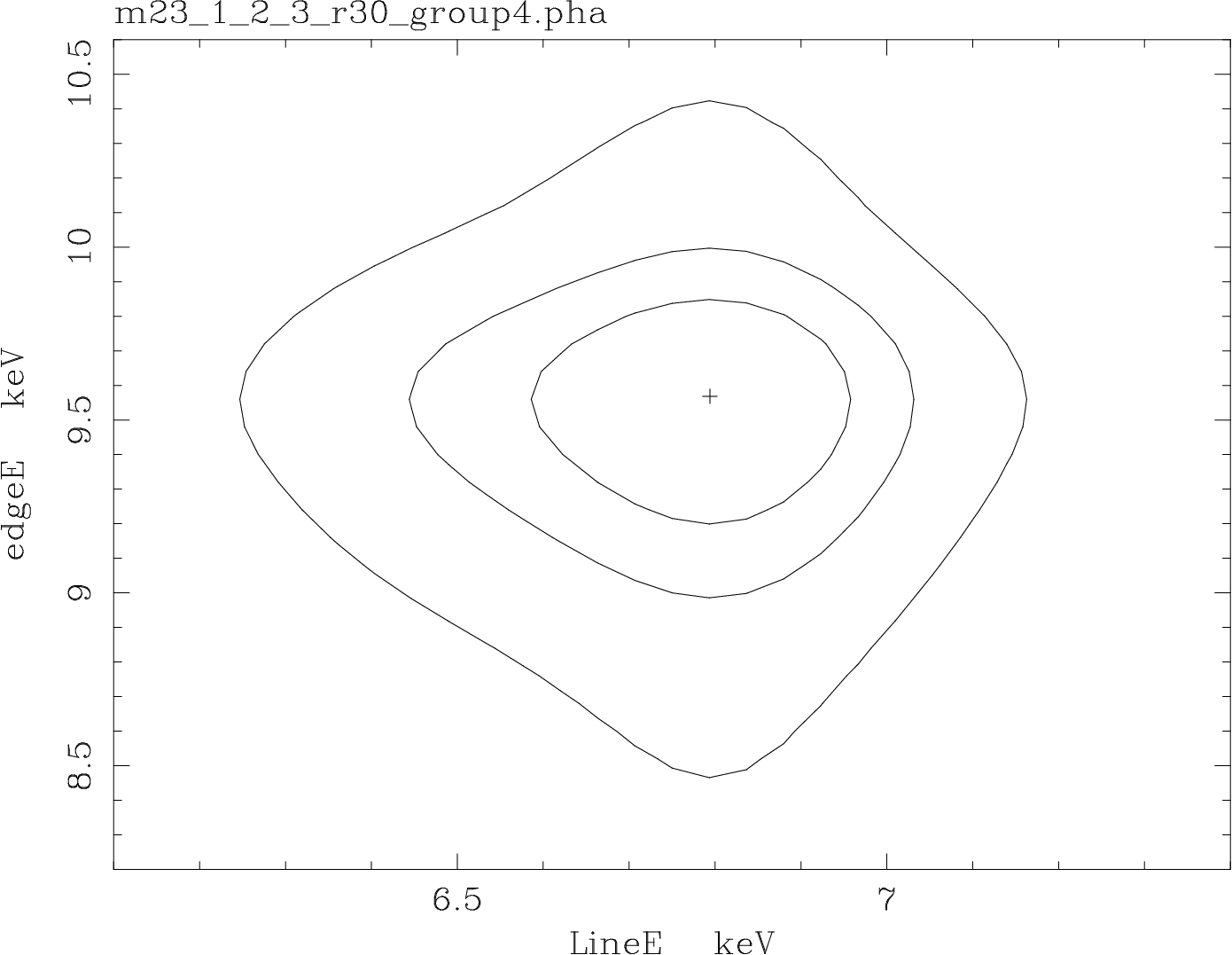, height=7.cm, angle=-90, silent=}
\caption{The 68, 90 and 99 \% confidence contours of line versus edge energy.}
\end{figure}

\subsection {The 0.1--10 keV spectrum}

The LECS spectrum was rebinned and grouped in the 0.1--4 keV 
energy range following the same criteria adopted for the MECS data.
The LECS and MECS spectra were fitted simultaneously leaving the 
relative normalization free to vary in the range 0.7--1.0 
(Fiore et al. 1999) to take into account
the intercalibration systematics between the two instruments.
The Galactic column density towards Ark 564,  
$N_{H Gal} = 6.4 \times 10^{20}$ cm$^{-2}$ (Dickey \& Lockman 1990),  
is obtained from 21 cm radio measurements averaged over relatively 
large portions of the sky (several tens of arcmin$^2$). 
As a result, $N_H$ fluctuations on smaller scales could introduce errors
of the order of 10$^{20}$ cm$^{-2}$ (see Elvis et al. 1994).
The low--energy cold absorption column density, included in all fits, 
is free to vary, unless otherwise specified, 
to account for these uncertainties.

The origin of the absorption/emission features in the 6--10 keV energy range
(Sect.~4.1) was tested against a warm absorber and an ionized reflection
model ({\tt ABSORI} and {\tt PEXRIV} respectively in {\tt XSPEC}; Magdziarz \&
Zdziarski 1995), 
including a narrow K$\alpha$ emission line, over the
entire BeppoSAX energy range (Table 4). 
In both fits the temperature of the warm/hot gas is fixed at 10$^6$ K.
The free parameters  were the power law slope $\Gamma$, 
the line energy and equivalent width, the warm gas column density
$N_{H warm}$ (or the relative normalization of the reflected component
with respect to the incident continuum $R$) and
the ionization parameter $\xi$ (defined as $\frac{L}{n r^2}$, where $L$ is the
luminosity of the source, $n$ the density of the medium and $r$ the
distance of the absorbing matter from the nucleus). 
Even though the overall shape of the BeppoSAX spectrum is relatively
well reproduced with the parameters shown in Table 4, 
and the low energy absorption column density ($\simeq$ 7 $\times$ 
10$^{20}$ cm$^{-2}$) 
is consistent with the Galactic value, neither
of the models provide an acceptable fit, leaving systematic residuals
in the data--to--model ratio especially at low energies, where a line--like
feature around 1 keV is clearly visible (Fig.~5). 
At high energies the best--fit values 
for the energy and EW of the iron K$\alpha$ line are consistent
with those obtained by fitting the 2--10 keV spectrum alone (see Table~3).
A single power law absorbed by an ionized medium can barely account 
for the K$\alpha$ line intensity.
Combining ionized reflection and warm absorption, the improvement in the 
overall fit quality is only marginal and the residuals are similar to those 
plotted in Fig.~5.  

\begin{figure}
\epsfig{file=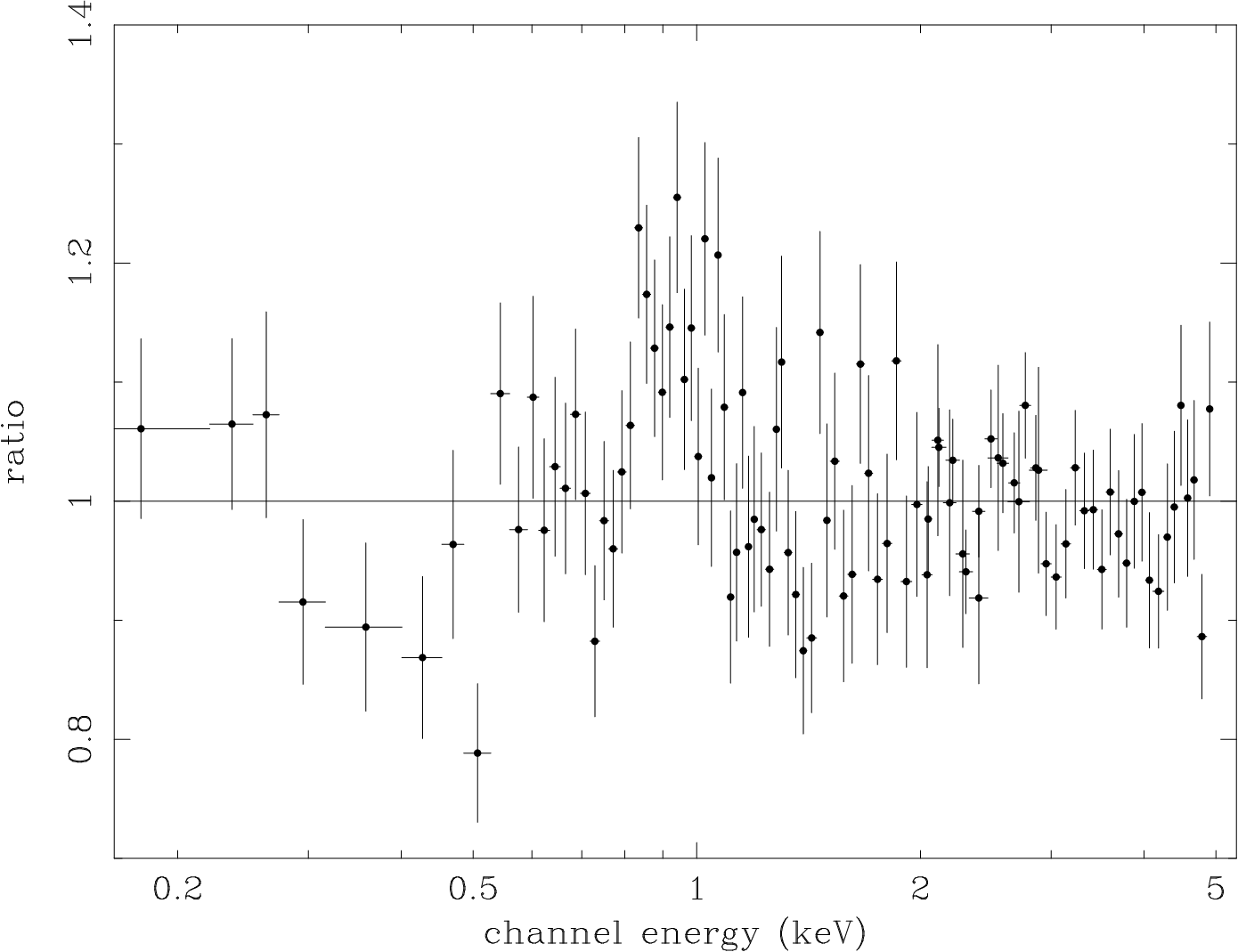, height=7.cm, angle=-90, silent=}
\caption{0.1-5 keV ratio with respect to a power law plus warm absorption
and reflection.}
\end{figure}

The data--to--model ratio of the BeppoSAX 0.1--10 keV spectrum 
obtained by fitting the 2--10 keV data and then extrapolating 
the power law down to 0.1 keV is shown in Fig.~6.
A strong soft component emerging below 1--2 keV is clearly evident.

\begin{figure}
\epsfig{file=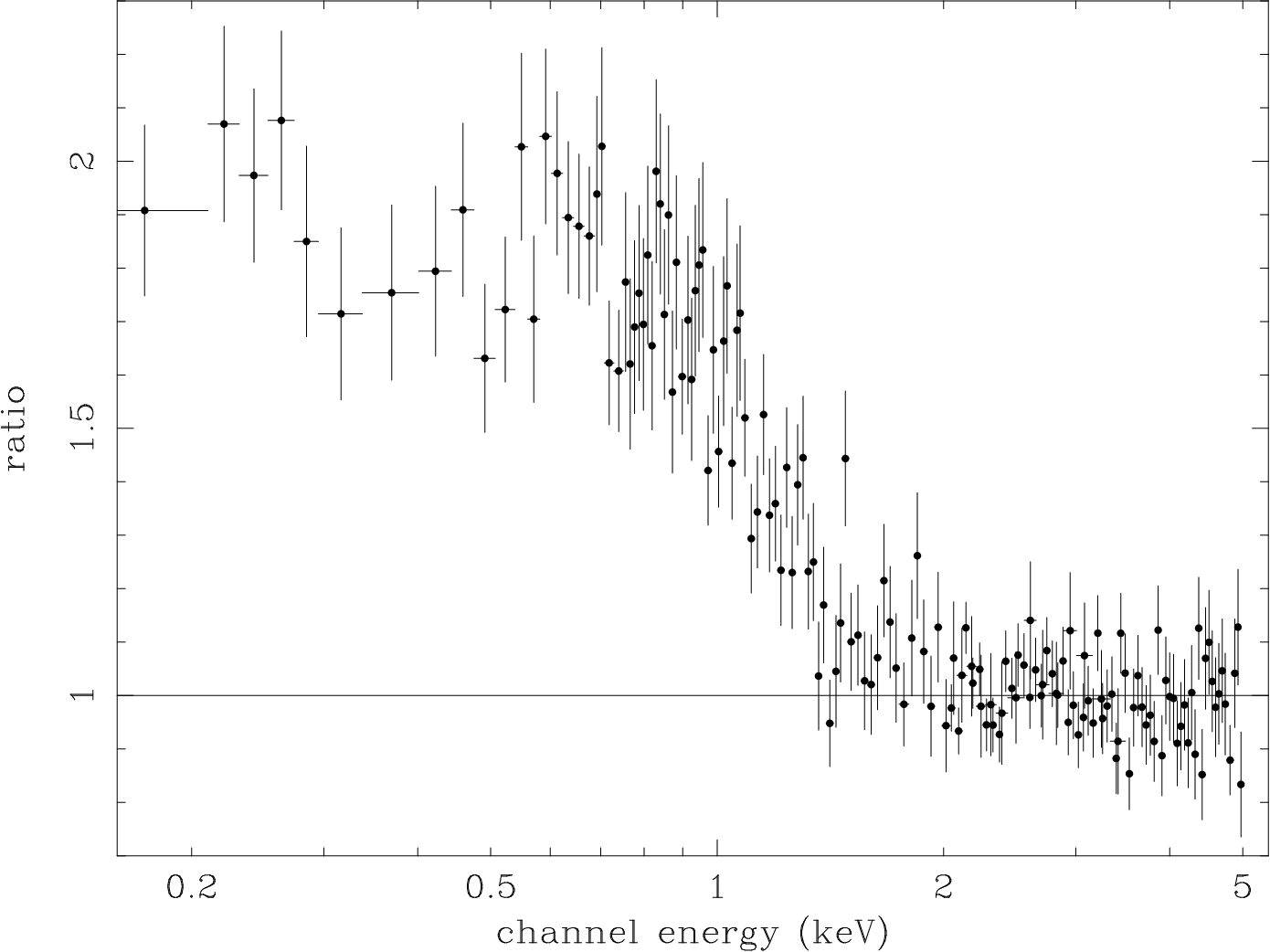, height=6.cm, angle=-90, silent=}
\caption{The data to model ratio with respect to a power law 
fit to the 2--10 keV spectrum. The residuals at energies greater 
than 5 keV are not reported for clarity.}
\end{figure}

\begin{table*}
\caption{Ionized reflection and absorption models}

\vglue0.3truecm
{\hfill\begin{tabular}{lcccccc}
 model & $\Gamma$ & $E_{line}$ (keV) & EW (eV) & $\xi$ (erg s$^{-1}$ cm) 
& $N_H$/$R$ & 
$\chi^2$/dof \\
\hline
 ABSORI+PO+LINE & 2.68$\pm$0.03 & 6.81$^{+0.16}_{-0.20}$ & 181$\pm$69 &
1667$^{+591}_{-368}$  & 4.3$^{+0.8}_{-0.6} \times 10^{22}$ &
305.6/260 \\
&&\\
 PEXRIV+PO+LINE & 2.61$\pm$0.03 & 6.76$^{+0.25}_{-0.37}$ & 87$\pm$65 &
 8776$^{+4959}_{-2914}$ & 0.7$\pm$0.1 & 363.5/260 \\
\hline

\end{tabular}\hfill}

\end{table*}

Spectral fits were next performed with two component models (Table~5).
The high--energy tail of a blackbody or a multicolour accretion 
disc spectrum plus a power law above $\sim$ 2 keV provide a 
good fit to the 0.1--10 keV X--ray continuum. 
The blackbody temperatures are of the order of 150--200 eV, while
the power--law photon index is about 2.4.
Although an acceptable statistical description of the data
is obtained if the soft X-ray spectrum is modelled with an
optically thin thermal plasma (the {\tt MEKAL} code in {\tt XSPEC}), 
the best fit abundances ($Z/Z_{\odot}$  $<$ 0.02) and the soft X-ray
variability are difficult to explain with such a model.
Alternative two component models for the 0.1--10 keV continuum,
such as a double power law or a broken power law, do not provide
an acceptable description of the data.

The derived best--fit value for the low--energy absorption column density is 
always lower than the Galactic value, suggesting the 
presence of an additional softer component.
Indeed, when the column density is kept fixed at the Galactic value,
systematic positive residuals with respect to the blackbody plus 
power--law fit, are present below $\sim$ 0.25 keV (the carbon band).
Not surprisingly, the addition of an extremely soft blackbody 
component ($kT \simeq$ 30 eV) improves the fit.

Given that the discrepancy between the column density values 
derived from the X--ray data and the 21 cm measurements ($\leq$ 
10$^{20}$ cm$^{-2}$) 
is comparable with the errors associated with the latter, 
the presence of an ultrasoft component should be considered with caution.
Finally even though the shape of the residuals in the 0.3--0.5 keV 
energy range (Fig.~6) are suggestive of an absorption edge,  
the size of the instrumental systematic errors at these energies 
(Orr et al. 1998) does not permit a search for the presence of an additional, 
possibly ionized, absorber.    
The unabsorbed 0.1--2.0 keV flux of the blackbody plus power--law model 
is 9.4 $\times$ 10$^{-11}$ erg cm$^{-2}$ s$^{-1}$, corresponding to a
luminosity of 2.6 $\times$ 10$^{44}$ erg s$^{-1}$.

\begin{table*}
\caption{LECS+MECS joint fits to the 0.1--10 keV continuum}

\vglue0.3truecm
{\hfill\begin{tabular}{l l l l l l}
  model & $N_H$ (10$^{20}$ cm$^{-2}$) & $kT_h$ (eV) & $\Gamma$ &  $\chi^2$/d.o.f. \\
\hline
 BB+PO & 5.2$\pm$0.2  & 159$\pm$7   & 2.40$\pm$0.03  & 272.6/262 \\
&&\\
 DISKBB+PO & 5.6$\pm$0.3  & 210$\pm$12  & 2.39$\pm$0.04  & 279.9/262 \\
&&\\
 MEKAL+PO  & 5.9$\pm$0.3 & 380$^{+44}_{-37}$ & 2.37$\pm$0.05  & 283/261 \\
&&\\
\hline

\end{tabular}\hfill}


\end{table*}

\section{Comparison with ASCA results} 

The line--like feature around 1 keV, originally discovered by ASCA, 
was successfully modelled by a power law continuum plus a Gaussian emission 
line (TGN99) or by an ionized reflection model plus an {\tt O VIII} 
recombination edge at 0.87 keV (V99).
In order to compare the BeppoSAX data with the ASCA observations, 
the same models were fitted to the LECS plus MECS spectra. 

The power law plus Gaussian line model returns an acceptable fit only
for very high values of the line width ($\sigma \simeq$ 0.6 or even larger), 
leaving the line energy basically unconstrained.
The profile of such a broad Gaussian line is equivalent to a blackbody 
continuum absorbed at low energy by the Galactic column density.

The addition of an {\tt O VIII} recombination edge 
(model {\tt REDGE} in {\tt XSPEC}) improves the statistical quality 
of the power--law plus ionized reflection fit. 
The best--fit values for the plasma temperature ($\sim$ 9 $\times$ 10$^5$ K
which is consistent with the assumed disc temperature of 10$^6$ K) and the  
{\tt O VIII} recombination feature EW ($\sim$ 90 eV) are in agreement 
with the V99 results. However, systematic residuals are left 
below 0.8 keV, suggesting that the recombination feature is too narrow to 
fully account for the shape of the soft X--ray spectrum.

We conclude that a single power law modified by warm/hot absorption and 
reflection features does not adequately represent the 
broad--band spectrum, and especially the soft X--ray spectral shape,
of Ark~564 as observed by BeppoSAX.

The discrepancy with the ASCA results is likely to be due
to both the different energy resolutions and spectral coverages 
of the detectors onboard ASCA and BeppoSAX.
In order to check whether the relatively weak emission line 
detected by ASCA could have been missed by the BeppoSAX LECS, 
a simulated 100 ks BeppoSAX observation of Ark~564 
was produced with the best--fit ASCA parameters quoted by TGN99.
We found that the simulated spectrum could be described well by a double 
blackbody plus power--law model, indicating that weak soft X--ray emission 
features can be only marginally studied with BeppoSAX.
On the other hand, the lack of sensitivity of the ASCA instruments below 
$\sim$ 0.6 keV coupled with the calibration uncertainties at these energies
prevents a detailed modelling of the soft X--ray continuum and 
of the strength of any emission/absorption feature.
An emission line at $\sim$ 1 keV is still statistically required 
if the ASCA spectrum is fitted with a blackbody plus power law model; 
however, the derived line EW is lower than that
obtained without a blackbody component (see Sect.~6). 

A further discrepancy between the BeppoSAX and ASCA results concerns the 
best--fit energy of the iron emission line, which is consistent with 
a neutral origin in the ASCA data. Moreover, while TGN99 find evidence
for a strong (EW $\sim$ 500 eV) line with a broad and asymmetric profile,
in the V99 analysis the emission feature is best fitted with 
a much weaker (EW $\sim$ 100 eV) and narrow Gaussian.
It is plausible that the different energy resolution and sensitivities 
of the detectors onboard ASCA and BeppoSAX coupled with the 
different prescriptions for the underlying continuum might account for 
part of this discrepancy.

\section{Discussion}

The X--ray spectral properties of Ark~564 as observed by BeppoSAX 
fit fairly well with the leading hypothesis of a higher accretion 
rate, relative to the Eddington value, and of a smaller black hole mass 
in NLS1 with respect to broad--line Seyfert 1 nuclei  
(e.g. Pounds et al. 1995; Boller et al. 1996; 
Laor et al. 1997). 
The steep soft excess, best fitted with the high--energy tail of a hot 
thermal component, is explained by a shift of the accretion disc spectrum
in the soft X--ray band. The strong flux of soft photons could lead to a 
strong Compton cooling of the coronal electrons and thus to a steep hard tail.
In some models the disc surface layers become strongly ionized when the 
accretion rate approaches the Eddington limit which fits nicely 
with the detection of ionized iron features. 
 
The reprocessing features originating in a highly ionized disc have been 
studied by Matt et al. (1993, 1996) and, more recently, by Ross et al. (1999) 
and Nayakshin et al. (2000). 
The intensities of the iron K$\alpha$ features as measured by BeppoSAX are 
in relatively good agreement with the model predictions for an ionization 
parameter $\xi \sim$ 10$^4$ erg cm s$^{-1}$ (compare Table~4). 
The best--fit values for the 
K$\alpha$ edge energy and optical depth are also consistent, within the 
errors, with the ASCA plus RXTE observation (V99), lending further support 
to the ionized reprocessor scenario. 
As discussed by Ross et al. (1999), the {\tt PEXRIV} model in {\tt XSPEC}
provides only a first approximation to the spectral shape of an ionized 
reflector. Unfortunately, the quality of the present data and the lack of 
MECS response above 10 keV does not allow testing of more detailed models. 

The increased reflectivity of the ionized accretion disk at low energies
cannot completely account for the strong soft excess in the BeppoSAX
data. An important consequence of the need for primary soft excess emission
is that the intensity and the origin 
of the 1 keV soft X--ray feature discovered by ASCA must probably be  
reconsidered. Even though there is no convincing evidence of such a feature 
in the BeppoSAX data, we cannot completely rule out this possibility.
We argue that the soft X--ray line intensity cannot be as strong as  
claimed by TGN99 and V99. Indeed, when fitting the ASCA data with a blackbody 
plus power law model, a 1 keV line is still required. The model dependent
EW (20--50 eV) is lower than the best fit value reported by TGN99 of
$\sim$ 70 eV and more similar to the best fit value of 29 eV reported by 
Leighly (1999b) from an independent analysis of the same ASCA data. 
As discussed by TGN99, a strong soft X--ray line can be only marginally 
reproduced
by emission from a warm/hot gas. Emission models computed for a wide range 
of column densities, ionization states, covering factors and iron abundances 
indicate that the largest EWs are of the order of 20--30 eV.
These results highlight the importance of a robust measurement of the 
underlying continuum shape for a better understanding of the origin 
of soft X--ray features in NLS1.
In conclusion, the soft X--ray spectrum of Ark~564 remains far from 
being unambiguously and convincingly explained. 
Better quality spectra, which will be obtained by the already
scheduled {\it Chandra} and {\it XMM--Newton} observations, 
are needed to settle this still controversial issue.

\par
The non--simultaneous optical, UV and X--ray spectral energy distribution 
(SED) is shown in Fig.~7.  
The UV data were retrieved from the HST archive using 
{\tt XSTARVIEW} and have been calibrated by the standard pipeline.
They agree relatively well with  those presented in Crenshaw et al. (1999) 
and with the blueward extrapolation of the optical data. A more detailed 
discussion of the full UV data set is beyond the scope of this 
paper. 
The flattening of the steep soft X--ray spectrum towards the lowest accessible
X--ray energies suggests that the energy density peaks in the soft X--ray band 
with a behaviour similar to that observed in RE J1034+393 
(Puchnarewicz et al. 2000). The lack of variability and the 
simultaneity of the RE J1034+393  optical--UV and X--ray data 
allowed spectral fits with accretion disc models.
The preferred interpretation was a nearly edge--on accretion
disc accreting at 0.3--0.7 L$_{Edd}$ onto a low mass black hole
(M $\sim$ a few 10$^6$ M$_{\odot}$).
It is likely that a similar set of parameters for the black hole mass 
and accretion rate are responsible for the Ark~564  
optical to X--ray spectrum. However, given the large--amplitude 
variability and the lack of simultaneous multifrequency observations,  
no attempt has been made to fit the observed SED with a disk model.

It is interesting to compare the Ark~564 and RE J1034+393 SED   
with that of Ton~S~180 (paper I): the 3 NLS1 
in the BeppoSAX core programme for which optical data are available.
The Ton~S~180 optical to X--ray spectrum (Fig.~5 in paper I) 
peaks somewhere at longer wavelengths; as a consequence, the
bulk of the energy density is emitted in the optical--UV band.
Indeed the shape of the soft--excess component 
of Ton~S~180 is best fitted by a steep power law while the soft X--ray 
components in Ark~564 and RE J1034+393 show a low--energy downwards curvature 
characteristic of thermal blackbody--like spectra. 

On the other hand, the 3 objects have similar high energy   
properties such as steep 2--10 keV spectra and ionized iron features
(though in RE J 1034+393 only a marginal detection has been reported; 
Pounds et al. 1995) suggesting that the high--energy 
spectrum is independent of the peak of the overall energy output. 
It is worth noting that the SED peak seems to be correlated with the
intensity of optical Fe~{\sc ii} multiplets which are particularly strong 
in Ton~S~180.
Further X--ray observations and coordinated optical--UV campaigns
are needed to clarify some of these issues.

\begin{figure}
\epsfig{file=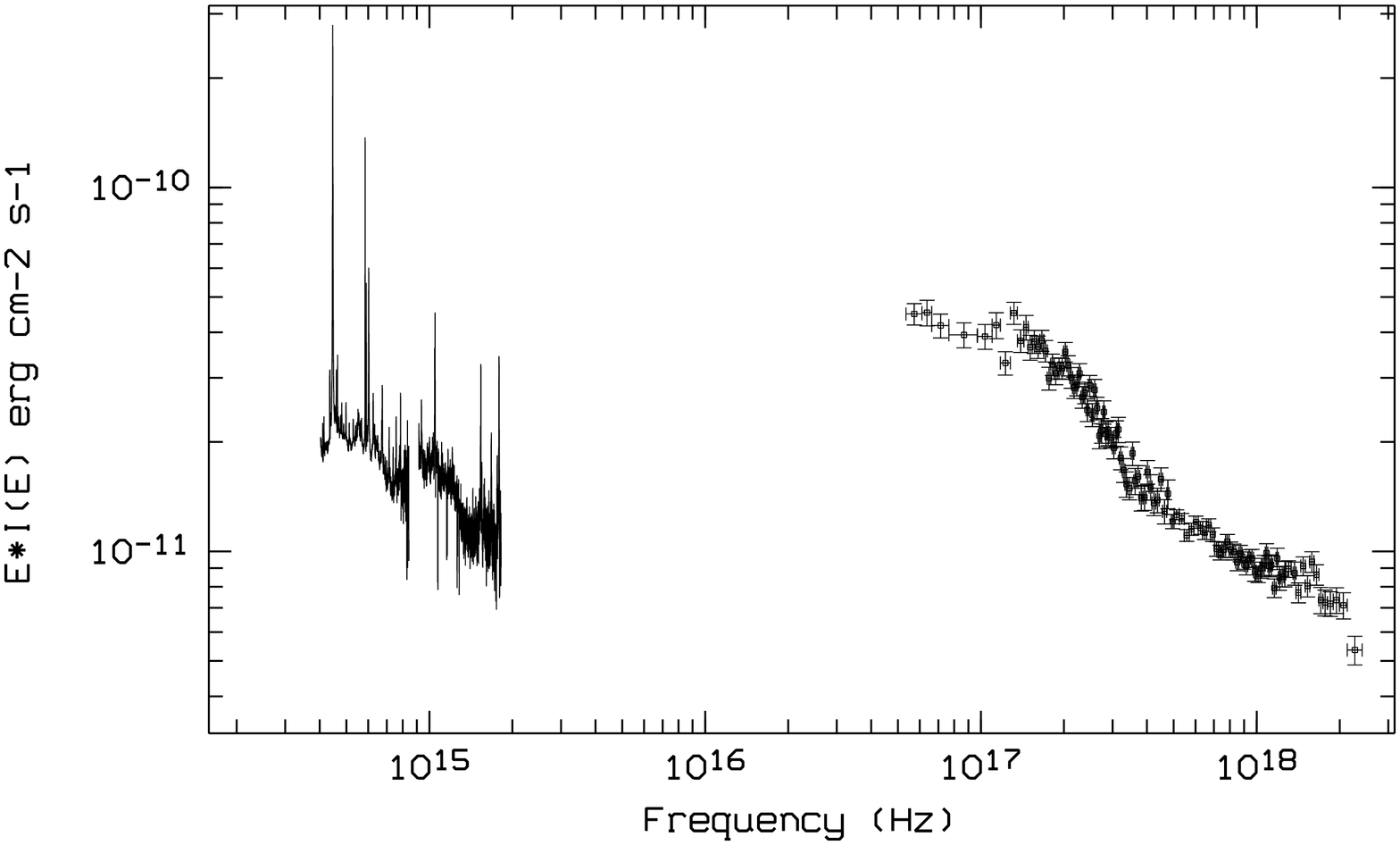, height=6.5cm, angle=-90, silent=}
\caption{Spectral energy distribution in the optical, UV and X--ray bands.}
\end{figure}

\section{Conclusions} 

We have analysed BeppoSAX data and a high-quality optical spectrum
of the bright Narrow-Line Seyfert~1 galaxy Ark~564.
Our main results are the following:

$\bullet$ The optical emission--line intensities and widths clearly confirm
the NLS1 nature of Ark~564. The Fe~{\sc ii} multiplets are not particularly
strong and are similar to those of normal Seyfert 1 galaxies.
  
$\bullet$ The evidence for soft X--ray spectral features is marginal
if the strong and possibly complex soft X--ray excess is modelled
with a blackbody or a multicolour accretion disc spectrum.

$\bullet$ The spectrum hardens at high energies though the power law 
photon index is rather steep and significantly steeper than 
the average value measured in broad line Seyfert 1 galaxies.

$\bullet$ The high--energy iron K$\alpha$ spectral features 
(the $\sim$ 6.8 keV line and the $\sim$ 9.5 keV edge)
strongly suggest reflection from a highly ionized, optically 
thick accretion disc.

$\bullet$ The optical to X--ray spectral energy distribution peaks
in the soft X--ray band with a behaviour similar to that observed in 
RE J1034+393. A rapidly accreting, low mass black hole
provides the most likely interpretation of the broad band observations.

\begin{acknowledgements}
We thank all the people who, at all levels, have made the SAX mission possible.
This research made use of SAXDAS linearized and cleaned event
files (Rev.2.0) produced at the BeppoSAX Science Data Center. 
Partial support from the Italian Space Agency under
the contract ASI--ARS--98--119, by the Italian Ministry for University
and Research (MURST) under grant Cofin--98--02--32 and by the NASA
Long Term Space Astrophysics grants NAG5--8107 and NAG5--7971 
is acknowledged.

\end{acknowledgements}


\begin{thebibliography}{}

\bibitem{d0} Boella G., Butler R.C., Perola G.C., 
et al., 1997a, A\&AS 122, 299

\bibitem{d0} Boella G., Chiappetti L., Conti G.,
et al., 1997b, A\&AS 122, 327

\bibitem{d0} Boller T., Brandt W.N, Fink H.H., 1996, A\&A 305, 53

\bibitem{d0} Boller T., Brandt W.N, Fabian A.C., Fink H.H., 1997, MNRAS 289, 393 

\bibitem{d0} Brandt W.N., Fabian A.C., Nandra K., Reynolds C.S.,
Brinkmann W., 1994, MNRAS 271, 958 

\bibitem{d0} Brandt W.N., Mathur S.,  Elvis M., 1997, MNRAS 285, 25  

\bibitem{d0} Comastri A., Fiore F., Guainazzi M., et al. 1998, A\&A 333, 31

\bibitem{d0} Crenshaw D.M., Kraemer S.B., Boggess A., et al. 1999, ApJ 516, 750  
\bibitem{d0} Dickey J.M., Lockman F.J., 1990, ARA\&A 28, 215  

\bibitem{d0} Elvis M., Lockman F.J., Fassnacht C., 1994, ApJS 95, 413 

\bibitem{d0} Fiore F., Laor A., Elvis M., Nicastro F., Giallongo E., 1998, 
ApJ 503, 607 

\bibitem{d0} Fiore F., Guainazzi M., Grandi P., 1999, Handbook for
BeppoSAX NFI spectral analysis \\
ftp://www.sdc.asi.it/pub/sax/doc/software\_docs/saxabc\_v1.2.ps.gz

\bibitem{d0} Frontera F., Costa E., Dal Fiume D., et al. 1997, A\&AS 122, 357 

\bibitem{d0} Guainazzi M., Matteuzzi A., 1997 SAX/SDC Technical Report 

\bibitem{d0} Joly M., 1988, A\&A 192, 87

\bibitem{d0} Laor A., Fiore F., Elvis M., et al., 1997, ApJ 477, 93 

\bibitem{d0} Leighly K.M., 1999a, ApJS 125, 297

\bibitem{d0} Leighly K.M., 1999b, ApJS 125, 317 

\bibitem{d0} Magdziarz P., Zdziarski A.A., 1995, MNRAS 273, 837 

\bibitem{d0} Manzo G., Giarrusso S., Santangelo A. et al. 1997, 
A\&AS 122, 341

\bibitem{d0} Matt G., Fabian A.C.,  Ross R.R., 1993, MNRAS 262, 179 

\bibitem{d0} Matt G., Fabian A.C.,  Ross R.R., 1996, MNRAS 278, 1111

\bibitem{d0} Nayakshin S., Kazanas D., Kallman T.R., 2000, ApJ 537, 833

\bibitem{d0} Orr A., Parmar A.N., Yaqoob T., Guainazzi M., 1998  Nuclear Physics B (Proc. Suppl.) 69/1--3, 496   

\bibitem{d0} Parmar, A.N., Martin D.D.E., Bavdaz M., et al. 1997, 
A\&AS 122, 309 

\bibitem{d0} Pounds K.A., Done C., Osborne J.P, 1995, MNRAS 277, L5  

\bibitem{d0} Puchnarewicz E.M., Mason K.O., Siemiginowska A., et al. 2000, ApJ in press   

\bibitem{d0} Ross, R.R., Fabian A.C., Young A.J., 1999,  MNRAS 306, 461 

\bibitem{d0} Turner T.J., George I.M., Nandra K., 1998, ApJ 508, 648

\bibitem{d0} Turner T.J., George I.M., Nandra K., Turcan D., 1999a, 
ApJ 524, 667

\bibitem{d0} Turner T.J., George I.M., Netzer H. 1999b, ApJ 526, 52 (TGN99)


\bibitem{d0} Vaughan S., Reeves J., Warwick R., Edelson R. 1999a, MNRAS
309, 113

\bibitem{d0} Vaughan S., Pounds K.A., Reeves J., Warwick R.,
 Edelson R. 1999b, MNRAS 308, L34 (V99)

\end{thebibliography}
\end{document}